\documentclass[runningheads]{llncs}
\usepackage[T1]{fontenc}
\usepackage{graphicx,wrapfig,lipsum,booktabs}
\usepackage{amsmath}
\usepackage{amssymb}
\usepackage{booktabs}
\usepackage{csquotes}
\usepackage{multirow}
\usepackage{mathtools}
\usepackage{makecell}
\usepackage[normalem]{ulem}
\usepackage[pagebackref,breaklinks,colorlinks]{hyperref}
\usepackage[usenames,dvipsnames,svgnames,table]{xcolor}
\usepackage{subcaption}
\usepackage{xspace}
\usepackage{amssymb, amsmath}
\usepackage{dsfont}
\usepackage{xcolor}

\makeatletter
\DeclareRobustCommand\onedot{\futurelet\@let@token\@onedot}
\def\@onedot{\ifx\@let@token.\else.\null\fi\xspace}
\def\eg{\emph{e.g}\onedot} 

\def\ie{\emph{i.e}\onedot}

\def\etal{\emph{et al}\onedot}
\makeatother

\renewcommand{\vec}[1]{{\mathbf #1}}
\usepackage{pifont}
\usepackage{hyperref}
\hypersetup{
    colorlinks=true,
    linkcolor=blue,
    filecolor=magenta,      
    urlcolor=cyan,
}
\usepackage[font=scriptsize]{caption}
\usepackage{mwe}
\usepackage[capitalize]{cleveref}
\crefname{section}{Sec.}{Secs.}
\Crefname{section}{Section}{Sections}
\Crefname{table}{Table}{Tables}

\begin{document}
\title{A Robust Volumetric Transformer for Accurate 3D Tumor Segmentation}
\titlerunning{VT-UNet}
\author{Himashi Peiris 
\inst{1}, Munawar Hayat 
\inst{3}, Zhaolin Chen 
\inst{1,2}, Gary Egan 
\inst{2}, Mehrtash Harandi 
\inst{1}}

\authorrunning{H. Peiris et al.}
%
\institute{Department of Electrical and Computer Systems Engineering, Faculty of Engineering, Monash University, Melbourne, Australia. \and Monash Biomedical Imaging (MBI), Monash University, Melbourne, Australia. \and  Department of Data Science \& AI, Faculty of IT, Monash University, Melbourne, Australia.\\
%
\email{\{Edirisinghe.Peiris, Munawar.Hayat, Zhaolin.Chen, Gary.Egan, Mehrtash.Harandi\}@monash.edu}}

\maketitle              
\vspace{-8mm}
\begin{abstract}
We propose a Transformer architecture for volumetric segmentation, a challenging task that requires keeping a complex balance in encoding local and global spatial cues, and preserving information along all axes of the volume. Encoder of the proposed design benefits from self-attention mechanism to simultaneously encode local and global cues, while the decoder employs a  parallel self and cross attention formulation to capture fine details for boundary refinement. Empirically, we show that the proposed design choices result in a computationally efficient model, with competitive  and  promising results on the Medical Segmentation Decathlon (MSD) brain tumor segmentation (BraTS) Task. We further show that the representations learned by our model are robust against data corruptions. \href{https://github.com/himashi92/VT-UNet}{Our code implementation is publicly available}.
\end{abstract}
\section{Introduction}
\label{sec:intro}
\begin{wrapfigure}{r}{4.2cm}
\scriptsize
\vspace{-8mm}
\centering
    \includegraphics[width=1\linewidth]{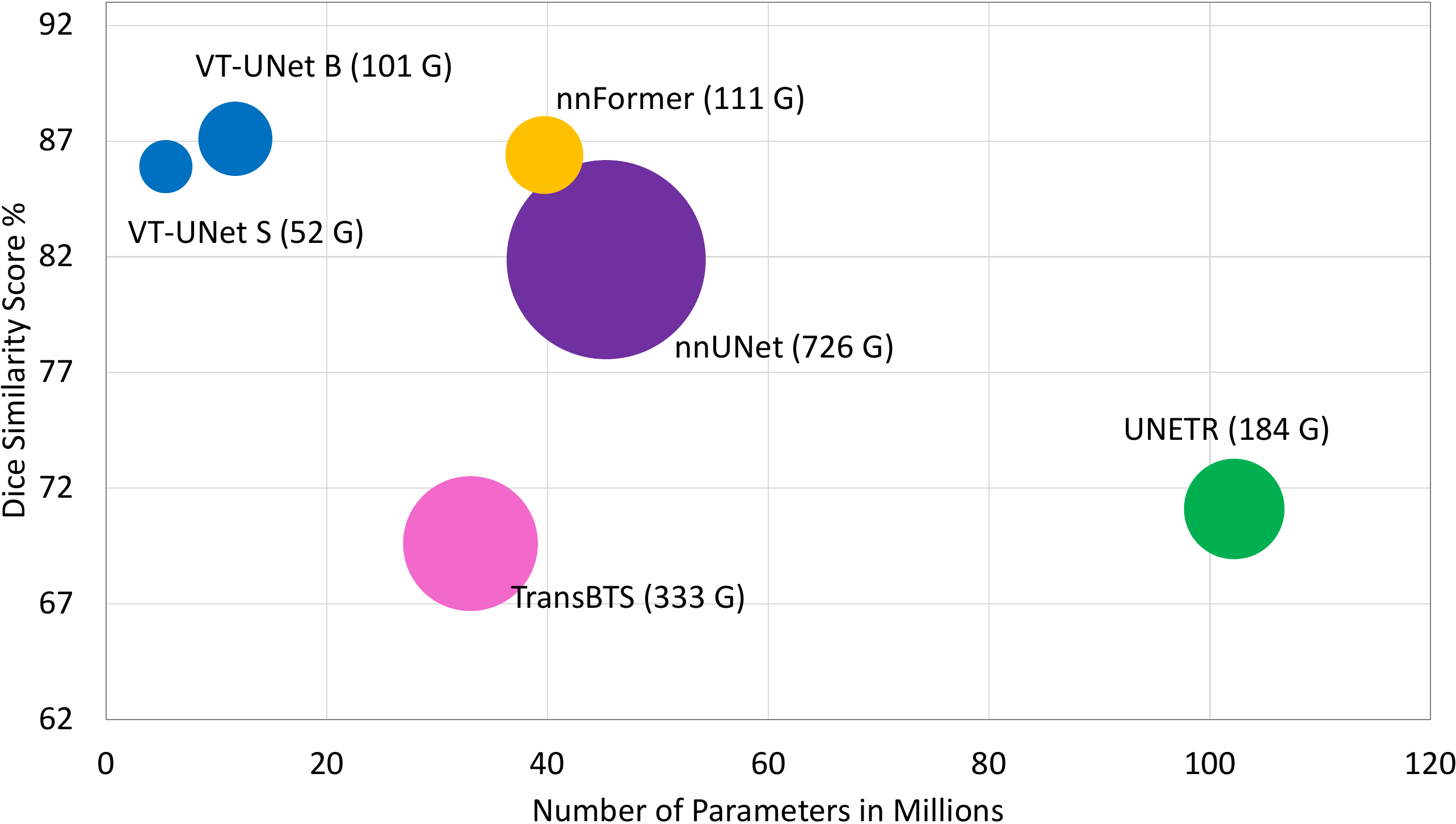}
    \caption{ The model size vs Dice Similarity Coefficient (DSC) is shown in this plot.  Circle size indicates Computational Complexity by FLOPs. VT-UNet achieves highest DSC compared to SOTA methods while maintaining a smaller model size and low computational complexity.
    }
    \label{fig:teaser}
    \vspace{-7mm}
\end{wrapfigure}
Inspired by the strong empirical results of the transformer based models in computer vision~\cite{dosovitskiy2020image,carion2020end,liu2021video}, their promising generalization and robustness characteristics~\cite{shao2021adversarial}, and their flexibility to model long range interactions, we propose a volumetric transformer architecture for segmentation of 3D medical image modalities (\eg, MRI, CT), called VT-UNet. Earlier efforts to develop transformer based segmentation models for 3D medical scans have been shown to outperform  state-of-the-art CNN counterparts~\cite{cao2021swin}. However, these methods divide 3D volumes into 2D slices and process 2D slices as inputs~\cite{cao2021swin,chen2021transunet}. As such, considerable and potentially critical volumetric information, essential to encapsulating inter-slice dependencies, is lost. While some hybrid approaches (using both convolutional blocks and Transformer layers) keep the 3D volumetric data intact~\cite{wang2021transbts,hatamizadeh2022unetr,zhou2021nnformer}, the design of purely transformer based architecture, capable of keeping intact the volumetric data at input, is yet unexplored in the literature. Our work takes the first step in this direction, and proposes a model, which not only achieves better segmentation performance, but also demonstrates better robustness against data artefacts. 
The Transformer models have highly dynamic and flexible receptive field and are able to capture long-range interactions, yet designing a Transformer based UNet architecture for volumetric segmentation remains a challenging task. This is because: \textbf{(1)} Encapsulating voxel information and capturing the connections between arbitrary positions in the volumetric sequence is not straightforward. Compared with Transformer based approaches for 2D image segmentation~\cite{cao2021swin}, the data in each slice of the volume is connected to three views and discarding either of them can be detrimental. \textbf{(2)} Preserving spatial information in a volume is a daunting task. Even for 2D images, while breaking the image into patches and projecting patches into tokens as introduced in \underline{Vi}sion \underline{T}ransformer (ViT), local structural cues can be lost, as shown in Tokens-to-token ViT~\cite{yuan2021tokens}. Effectively encoding the local cues while simultaneously capturing global interactions along multiple axes of a volume is therefore a challenging task. \textbf{(3)} Due to the  quadratic complexity of the self-attention, and large size of 3D volume tensor inputs, designing a Transformer based segmentation model, which is computationally efficient, requires careful design considerations.
Our proposed VT-UNet model effectively tackles the above design challenges by proposing a number of modules. In our UNet based architecture, we develop two types of Transformer blocks. First, our blocks in the encoder which directly work on the 3D volumes, in a hierarchical manner, to jointly capture the local and global information, similar in spirit to the Swin Transformer blocks~\cite{liu2021swin}. Secondly, for the decoder, we introduce parallel cross-attention and self-attention in the expansive path, which creates a bridge between queries from the decoder and keys \& values from the encoder. By this parallelization of the cross-attention and self-attention, we aim to preserve the full global context during the decoding process, which is important for the task of segmentation. 
Since VT-UNet is free from convolutions and combines attention outputs from two modules during the decoding, the order of the sequence is important to get accurate predictions. Inspired by~\cite{vaswani2017attention}, apart from applying relative positional encoding while computing attention in each Transformer block, we augment the decoding process and inject the complementary information extracted from Fourier feature positions of the tokens in the sequence. 
In summary, our major contributions are, \textbf{(1)} We reformulate volumetric tumor segmentation from a sequence-to-sequence perspective, and propose a UNet shaped \textit{Volumetric Transformer} for multi-modal medical image segmentation.
\textbf{(2)} We design an encoder block with two consecutive self attention layers to jointly capture local and global contextual cues. Further, we design a decoder block which enables parallel (shifted) window based self and cross attention. This parallelization uses one shared projection of the \emph{queries} and independently computes cross and self attention. To further enhance our features in the decoding, we propose a convex combination approach along with Fourier positional encoding. 
\textbf{(3)} Incorporating our proposed design choices, we substantially limit the model parameters while maintaining lower FLOPs compared to existing approaches (see~\cref{fig:teaser}). \textbf{(4)} We conduct extensive evaluations and show that our design achieves state-of-the-art volumetric segmentation results, alongwith enhanced robustness to data artefacts.

\section{Methodology}
\label{sec:method}
\vspace{-9mm}
\begin{figure*}[h]
\scriptsize
\centering
\begin{tabular}{{c@{ } c@{ } c@{ }}}
     \multirow{3}{*}[0.4in]{{\includegraphics[width=0.6\linewidth]{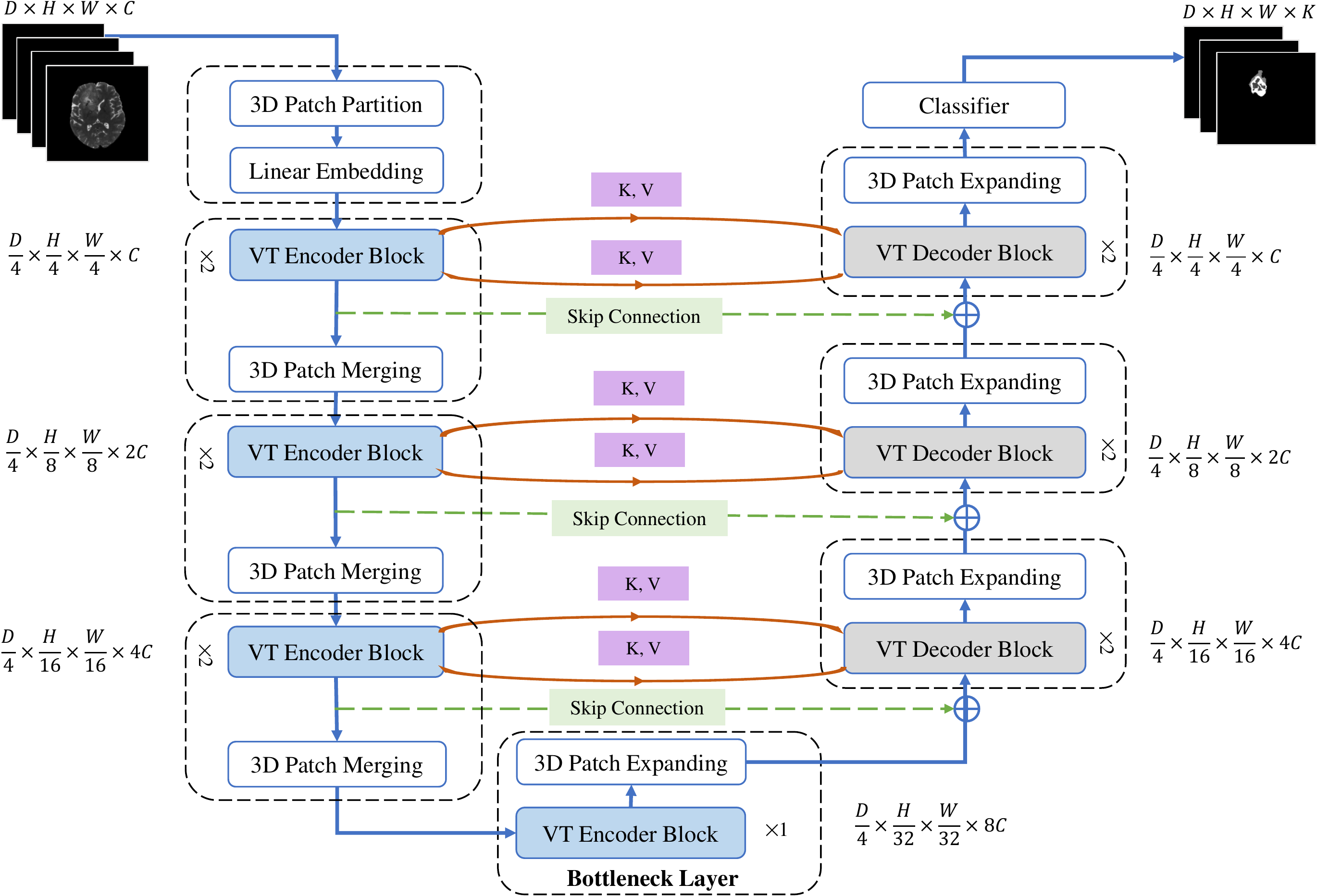}}} &
     \hspace{0.08cm} &
    {\includegraphics[width=0.3\linewidth]{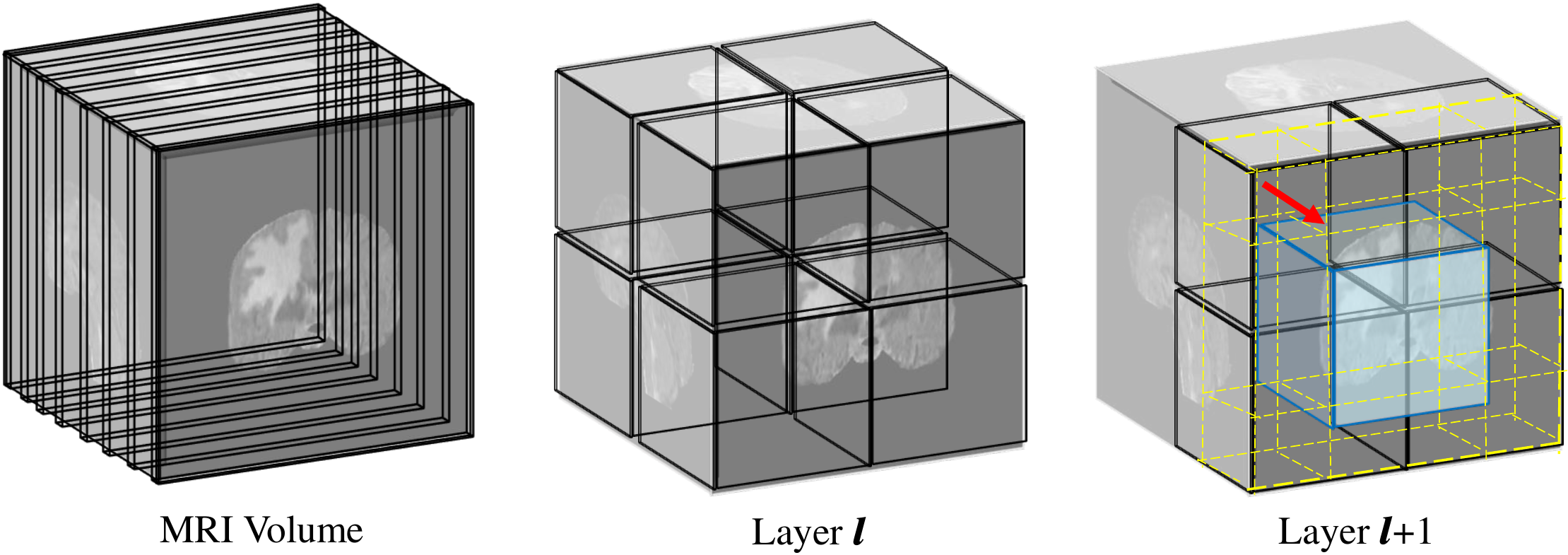}}\\
    & \hspace{0.08cm} & \footnotesize{(b)}\\
    & \hspace{0.08cm} & {\includegraphics[width=0.3\linewidth]{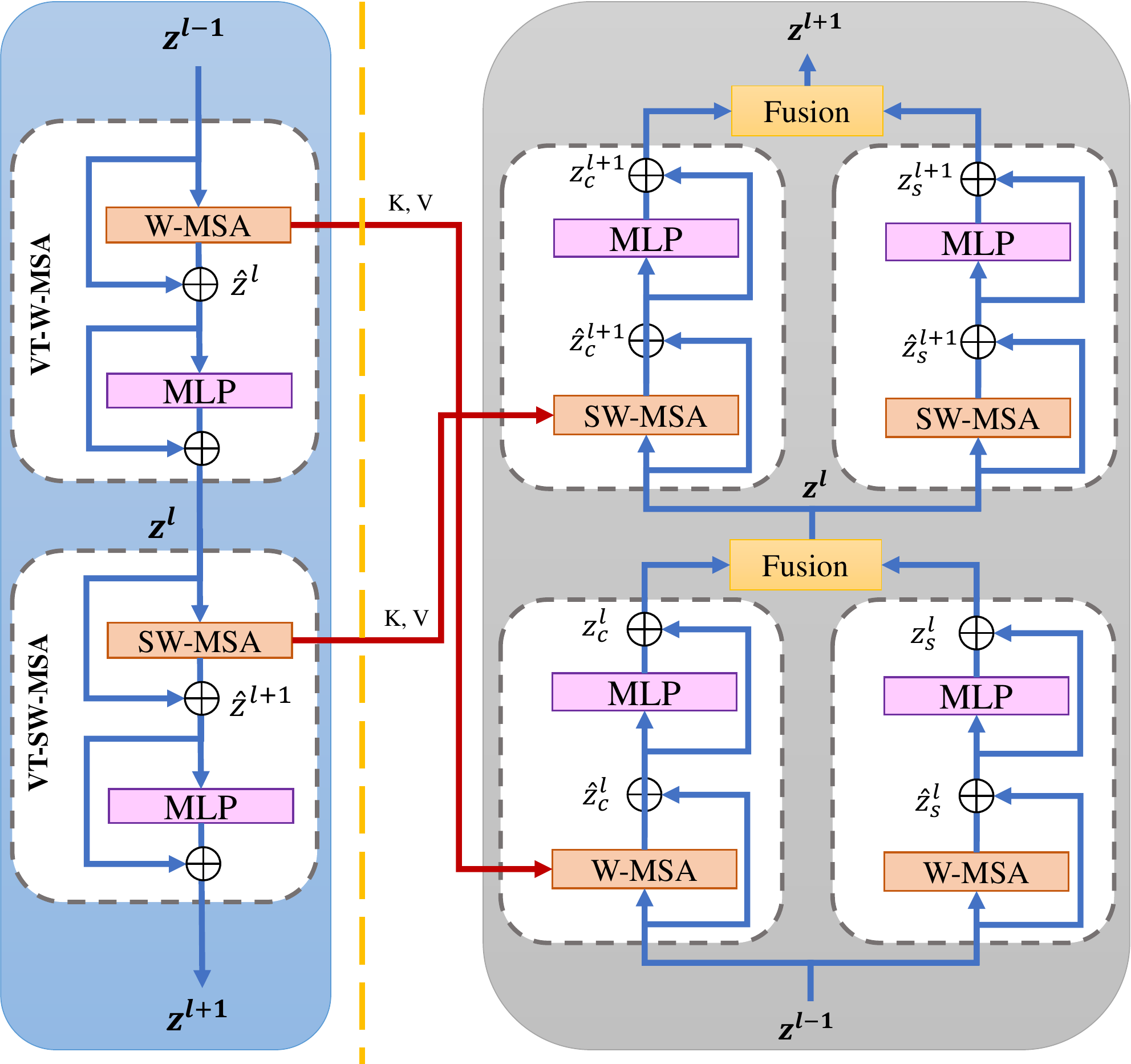}}\\
    \footnotesize{(a)} & \hspace{0.08cm} & \footnotesize{(c)}\\
    {\includegraphics[width=0.6\linewidth]{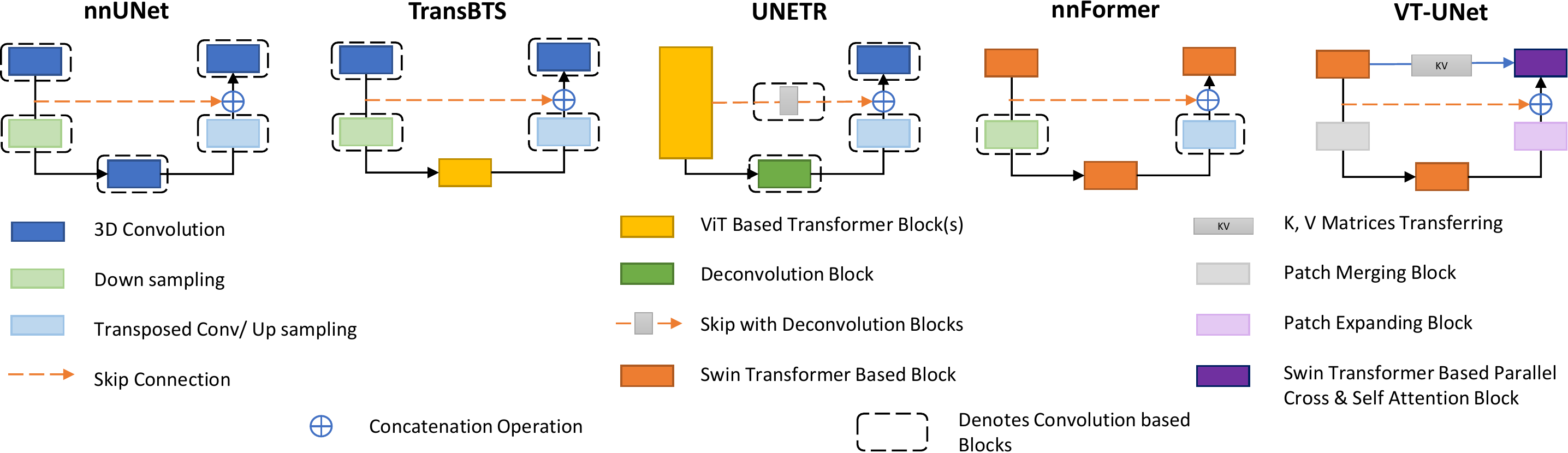}}
    & \hspace{0.08cm} & {\includegraphics[width=0.22\linewidth]{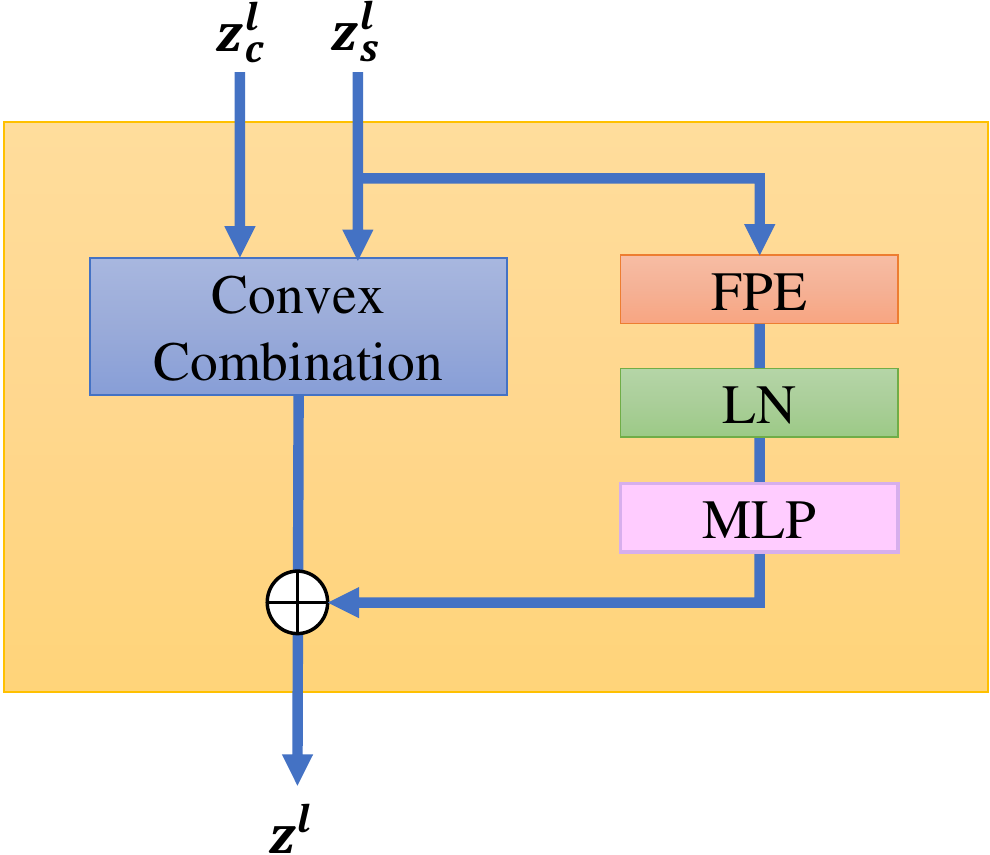}}\\
    \footnotesize{(d)} & \hspace{0.08cm} & \footnotesize{(e)}\\
  \end{tabular}
\caption{(a) Illustrates VT-UNet Architecture. Here, $k$ denotes the number of classes. (b) shows visualization of Volumetric Shifted Windows. Consider an MRI volume of size $D \times H \times W$ with $D = H = W = 8$ for the sake of illustration. Further, let the window size for partitioning the volume 
be $P \times M \times M$ with $P = M = 4$. Here, layer $\emph{l}$ adopts the regular window partition in the first step of \underline{V}olumetric \underline{T}ransformer(VT) block which results in $2 \times 2 \times 2 = 8$ windows. Inside layer $\emph{l} + 1$, volumetric windows are shifted by ($\frac{P}{2}$, $\frac{M}{2}$, $\frac{M}{2}$) =(2, 2, 2) tokens. This results in $3 \times 3 \times 3 = 27$  windows. (c) shows VT Encoder-Decoder Structure. (d)Encoder-Decoder structural comparison with other SOTA methods. The proposed VT-UNet architecture has no convolution modules and is purely based on Transformer blocks. (e) Illustrates the structure of the Fusion Module.}
\label{fig:architecture}
\end{figure*}
\vspace{-4mm}
We denote vectors and matrices in bold lower-case $\vec{x}$ and bold upper-case $\vec{X}$, respectively. 
Let $\mathds{X} = \{\vec{x}_1,\vec{x}_2,\cdots,\vec{x}_\tau\}, \vec{x}_i \in \mathbb{R}^C$ be a sequence representing a signal of interest (\eg, an MRI volume). We call each $\vec{x}_i$ a token. We assume that tokens, in their original form,  might not be optimal for defining the span. Therefore, in \underline{S}elf-\underline{A}ttention (SA), we define the span by learning a linear mapping from the input tokens. This we can show with $\mathbb{R}^{\tau \times C_v} \ni \vec{V} = \vec{X}\vec{W}_\vec{V}$, where we stack tokens $\vec{x}_i$s into the rows of $\vec{X}$ (\ie, $\vec{X}=[\vec{x}_1|\vec{x}_2|\cdots\vec{x}_\tau]^\top$). 
Following previous work by Hu \etal~\cite{hu2019local}, we use a slight modification of the self-attention~\cite{vaswani2017attention}(see~\cite{khan2021transformers}) in our task as follows:
\vspace{-2mm}
\begin{equation}
    \text{SA}(\vec{Q}, \vec{K}, \vec{V}) = 
    \text{SoftMax}\Big(\vec{Q}\vec{K}^\top  \mathbin{/} \sqrt{C}+\vec{B}\Big)\vec{V},
    \label{eqn:attn}
\end{equation}
\noindent where $\mathbb{R}^{\tau \times \tau} \ni \vec{B}$ is trainable and acts as a relative positional bias across tokens in the volume with $\vec{V} = \vec{X}\vec{W}_\vec{V}$, $\vec{K} = \vec{X}\vec{W}_\vec{K}$, and $\vec{Q} = \vec{X}\vec{W}_\vec{Q}$.  In practice, computing SA for multiple attention heads several times in parallel is called \underline{M}ulti-head \underline{S}elf-\underline{A}ttention (MSA).
~\cref{eqn:attn} is the basic building block of our \underline{V}olumetric \underline{T}ransformer \underline{W}indow based \underline{M}ulti-head \underline{S}elf-\underline{A}ttention (VT-W-MSA), and the \underline{V}olumetric \underline{T}ransformer \underline{S}hifted \underline{W}indow based \underline{M}ulti-head \underline{S}elf-\underline{A}ttention (VT-SW-MSA), discussed next. 

\noindent
\textbf{Overview of VT-UNet.} \cref{fig:architecture} shows the conceptual diagram of the proposed volumetric transformer network or \textbf{VT-UNet} for short. The input to our model is a 3D volume of size $D \times H \times W \times C$. The output is a $D \times H \times W \times K$ dimensional volume, representing the presence/absence of voxel-level class labels ($K$ is the number of classes). 
Below, we discuss architectural form of the VT-UNet modules and explain the functionality and rationals behind our design in detail. 

\noindent \textbf{The VT Encoder. } 
The VT encoder  consists of 3D Patch Partitioning layer together with Linear Embedding layer, and 3D Patch merging layer followed by two successive VT encoder blocks.

\noindent
\textcolor{purple}{\textbf{3D Patch Partitioning.}}
Transformer-based models work with a sequence of tokens. The very first block of VT-UNet accepts  a  $D \times H \times W \times C$ dimensional medical volume (\eg, MRI) and creates a set of tokens by splitting the 3D volume into non-overlapping 3D patches (see~\cref{fig:architecture} (b)). The size of partitioning kernel is $P \times M \times M$, resulting in describing the volume by $\tau = \lfloor D/P \rfloor \times \lfloor H/M \rfloor \times \lfloor W/M \rfloor$ tokens.  The 3D patch partitioning is followed by a linear embedding  to map each token with dimensionality $P \times M \times M$ to a $C$ dimensional vector. Typical values for $M$, $P$ and $C$ according to our experiments are 4, 4, and 72, respectively.

\noindent
\textcolor{purple}{\textbf{VT Encoder Block.}} 
In ViT, tokens carry significant spatial information due to the way they are constructed.  The importance of performing SA by windowing in ViT  has been shown in several recent studies, most notably in Swin~Transformer~\cite{liu2021swin}. Following a similar principal in 
the design of Swin~Transformers, albeit for volumetric data, we propose 3D windowing operations in our \underline{VT} \underline{Enc}oder \underline{Bl}oc\underline{k}s (\textbf{VT-Enc-Blks}). In particular, we propose two types of windowing, namely regular window  and shifted window, which we show by \textbf{VT-W-MSA} and \textbf{VT-SW-MSA} for simplicity, respectively. \cref{fig:architecture}b provides the design specifics of  \textbf{VT-W-MSA} and \textbf{VT-SW-MSA}, while \cref{fig:architecture} (b) illustrates the windowing operation.
Both VT-W-MSA and VT-SW-MSA employ  attention layers with windowing, followed by a 2-layer \underline{M}ulti \underline{L}ayer \underline{P}erceptron (MLP) with \underline{G}aussian \underline{E}rror \underline{L}inear \underline{U}nit (GELU) non-linearity in between. A \underline{L}ayer \underline{N}ormalization (LN) is applied before every MSA and MLP, and a residual connection is applied after each module. 
The windowing enables us to inject inductive bias in modeling long range dependencies between tokens. In both VT-W-MSA and VT-SW-MSA, attention across tokens within a window helps representation learning.
In the VT-W-MSA, we split the volume evenly into smaller non-overlapping windows as illustrated in~\cref{fig:architecture}~(b). Since tokens in adjacent windows cannot see each other with VT-W-MSA, we make use of a shifted window in VT-SW-MSA (see the right most panel~\cref{fig:architecture}~(b)) which bridges tokens in adjacent windows of VT-W-MSA. The windowing is inspired by the Swin~Transformer~\cite{liu2021swin} and can be understood as generalization to volumetric data. Note that the windowing operation in our work resembles ~\cite{liu2021video} that extends the benefits of windowing beyond images to videos.
Putting everything together, the VT-Enc-Blk realizes the following functionality:
\begin{align}
    &{{\hat{\bf{z}}}^{l}} = \text{VT-W-MSA}\left( {\text{LN}\left( {{{\bf{z}}^{l - 1}}} \right)} \right) + {\bf{z}}^{l - 1},\nonumber \qquad 
    &{{\hat{\bf{z}}}^{l+1}} = \text{VT-SW-MSA}\left( {\text{LN}\left( {{{\bf{z}}^{l}}} \right)} \right) + {\bf{z}}^{l}, \nonumber\\
    &{{\bf{z}}^l} = \text{MLP}\left( {\text{LN}\left( {{{\hat{\bf{z}}}^{l}}} \right)} \right) + {{\hat{\bf{z}}}^{l}}, \qquad 
    &{{\bf{z}}^{l+1}} = \text{MLP}\left( {\text{LN}\left( {{{\hat{\bf{z}}}^{l+1}}} \right)} \right) + {{\hat{\bf{z}}}^{l+1}}, 
    \label{eqn:vt_block}
\end{align}
where ${\hat{\bf{z}}}^l$ and ${\bf{z}}^l$ denote the output features of the VT-W-MSA module and the MLP module for block $l$, respectively. 

\noindent
\textcolor{purple}{\textbf{3D Patch Merging.}}
We make use of 3D patch merging blocks to generate feature hierarchies in the encoder of VT-UNet. Having such  hierarchies is essential to generate finer details in the output for the dense prediction tasks~\cite{liu2021swin,chen2017deeplab}. 

After every VT-Enc-Blk, we merge adjacent  tokens along the spatial axes in a non-overlapping manner to produce new tokens. In doing so, we first  concatenate features of each group of $2 \times 2$ neighboring tokens. The resulting vector is  projected via a linear mapping to a space where the channel dimensionality of the tokens is doubled (see \cref{fig:architecture}). The benefit of patch merging is not limited to feature hierarchies. The computational complexity of SA is quadratic in the number of tokens~\cite{liu2021swin,liu2021video}. As such, patch merging reduces the FLOPs count  of the VT-UNet by a factor of 16 after each VT-Enc-Blk. To give the reader a better idea and as we will discuss in \textsection\cref{sec:exp}, the tiny VT-UNet model uses only 6.7\% FLOPs in comparison to its fully volumetric CNN counterpart~\cite{milletari2016v} while achieving a similar performance (slightly better indeed)! Please note that the patch merging block is not used in the bottleneck stage. 

\noindent \textbf{The VT Decoder. } 
After bottleneck layer which consists of a VT-Enc-Blk together with 3D Patch Expanding layer, the VT decoder starts with successive \underline{VT} \underline{Dec}oder \underline{Bl}oc\underline{k}s (\textbf{VT-Dec-Blks}), 3D patch expanding layers and a classifier at the end to produce the final predictions. There are some fundamental design differences between VT-Enc-Blk and VT-Dec-Blk which we will discuss next.

\noindent
\textcolor{purple}{\textbf{3D Patch Expanding.}} 
This functionality is used to somehow revert the effect of patch merging. In other words and in order to construct the output with the same spatial-resolution as the input, we need to create new tokens in the decoder. For the sake of discussion, consider the patch expanding  after the bottleneck layer (see the middle part of \cref{fig:architecture}). The input tokens to the patch expanding are of dimensionality $8C$. In the patch expanding, we first increase the dimensionality of the input tokens by a factor of two using  a linear mapping. Following a reshaping, we can obtain $2 \times 2$ tokens with dimensionality $4C$ from the resulting vector of dimensionality $2 \times 8C$. This, we will reshape along the spatial axes and hence for $D/4 \times H/32 \times W/32 \times 8C$, we create \(D/4 \times H/16 \times W/16  \times 4C\) tokens. 

\noindent
\textcolor{purple}{\textbf{VT Decoder Block.}} The UNet~\cite{ronneberger2015u} and its variants~\cite{oktay2018attention,zhou2018unet++} make use of lateral connections between the encoder and the decoder to produce fine-detailed predictions. This is because the spatial information is lost, at the expense of attaining higher levels of semantics,  as the input passes through the encoder. The lateral connections in the UNet makes it possible to have the best of  both worlds,  spatial information from lower layers and semantic information from upper layers (along the computational graph). Having this in mind, we propose a hybrid form of SA at the decoder side  (see~\cref{fig:architecture}b for an illustration). Each VT-Dec-Blk receives the generated tokens of its previous VT-Dec-Blk along with the key ($\vec{K}_E$) and value ($\vec{V}_E$) tokens from the VT-Enc-Blk sitting at the same stage of VT-UNet, see~\cref{fig:architecture}a. Recall that a  VT-Enc-Blk has two SA blocks with regular and shifted windowing operations. VT-Dec-Blk enjoys similar windowing operations but makes use of four SA blocks grouped into SA module and  \underline{C}ross \underline{A}ttention(CA) module. The functionality can be described as:
\begin{align}
    &{\text{SA}_r} = \text{SA}(\vec{Q}_D,\vec{K}_D,\vec{V}_D), \qquad 
    &{\text{CA}_l} = \text{SA}(\vec{Q}_D,\vec{K}_E,\vec{V}_E).
        \label{eqn:ca_sa}
\end{align}
Here, \emph{r} and \emph{l}, denote right and left branches of the decoder module. The right branch of the SA acts on tokens generated by the previous VT-Dec-Blk according to~\cref{eqn:ca_sa}. We emphasize on the flow of information from the decoder by the subscript $D$ therein. The left branch of the CA, however, uses the queries generated by the decoder along with the keys and values obtained from the VT-Enc-Blk at the same level in the computation graph. The idea here is to use the basis spanned by the encoder (which is identified by values) along with keys to  benefit from spatial information harvested by the encoder. These blocks, also use the regular and shifted windowing to inject more inductive bias into  the model. Note that the values and keys from the SA with the same windowing operation should be combined, hence the criss-cross connection form in \cref{fig:architecture} (c). 
\begin{remark}
One may ask why values and keys are considered from the encoder. We indeed studied other possibilities such as employing queries and keys from the encoder while generating values by the decoder. Empirically, the form described in \cref{eqn:ca_sa} is observed to deliver better and more robust outcomes and hence our choice in VT-UNet\footnote{\scriptsize We empirically observed that employing  keys and values from the encoder in CA yields faster convergence of VT-UNet. This, we conjecture, is due to having extra connections from the decoder to encoder during the back-propagation which might facilitate gradient flow.}.
\end{remark}
\noindent
\textcolor{purple}{\textbf{Fusion Module.}}
As illustrated in~\cref{fig:architecture} (e), tokens generated from the CA module and MSA module are combined together and fed to the next VT-Dec-Blk, ${\bf{z}}^l$ is calculated using by a linear function as:
\vspace{-2mm}
\begin{align}
    \vec{z}^l = \alpha~ {{\hat{\bf{z}}}_{c}^{l}} + (1 - \alpha)~{{\hat{\bf{z}}}_{s}^{l}} + \mathcal{F}({{\hat{\bf{z}}}_{s}^{l}}),
    \label{eqn:combine_function}
\end{align}
where $\mathcal{F}(\cdot)$ denotes \underline{F}ourier Feature \underline{P}ositional \underline{E}ncoding (FPE) and $\alpha$ controls the contribution from each CA and MSA module. Aiming for simplicity, in fusing tokens generated by the CA  and MSA, we use a linear combination with $\alpha =  0.5$\footnote{\scriptsize Breaking the Symmetry: This results in a symmetry, meaning that swapping $\hat{\vec{z}}_{c}^{l}$ and $\hat{\vec{z}}_{s}^{l}$ does not change the output. To break this symmetry and also better  encapsulate  object-aware  representations that  are  critical  for  anatomical pixel-wise segmentation, we supplement the tokens generated from MSA by a the 3D FPE. The 3D FPE employs sine and cosine functions with different frequencies~\cite{vaswani2017attention} to yield a unique encoding scheme for each token. The main idea is to use a sine/cosine function with a  high frequency and modulate it across the dimensionality of the tokens while changing the frequency according to the location of the token within  the 3D volume.}.

\noindent
\textcolor{purple}{\textbf{Classifier Layer.}}
After the final 3D patch expanding layer in the decoder, we introduce a classifier layer which includes a 3D convolutional layer to map deep $C$ dimensional features to $K$ segmentation classes. 

\paragraph{A note on computational complexity.} 
The computational complexity of the SA described in \cref{eqn:attn} is dictated by computations required for obtaining $\vec{Q}, \vec{K}, \vec{V}$, computing $\vec{Q}\vec{K}^\top$ and obtaining the resulting tokens by applying the output of the Softmax (which is a $\tau \times \tau$ matrix) to $\vec{V}$. This adds up to $\mathcal{O}\big(3 \tau C^2 + 2\tau^2C\big)$, where $C$ and $\tau$ are the dimensionality and the number of tokens, respectively. Windowing will reduce the computational load of the SA according to $\mathcal{O}\big(3 \tau C^2 + 2\tau \kappa C\big)$ where we have assumed that  tokens are grouped into $\kappa$ windows and SA is applied within each window. In our problem, where tokens are generated from volumetric data, $\tau \gg \kappa$ and hence windowing not only helps in having better discriminatory power, but also it helps in reducing the computational load\footnote{\scriptsize For the sake of simplicity and explaining the key message, we have made several assumptions in our derivation. First, we have assumed $C_k=C_v=C$. We also did not include the FLOPs needed to compute the softmax. Also, in practice, one uses a multi-head SA, where the computation is break down across several parallel head working on lower dimensional spaces (\eg, on for $\vec{V}$, we use $C/h$ dimensional spaces where $h$ is the number of heads). This will reduce the computational load accordingly. That said, the general conclusion provided here is valid.}.

\section{Related Work}
\label{sec:literature}
Vision Transformers have shown superior empirical results for different computer vision tasks~\cite{touvron2021training,arnab2021vivit,liu2021video}, with promising characteristics. For example and  as compared with the CNNs, they are less biased towards texture~\cite{naseer2021intriguing}, and show better generalization and robustness~\cite{shao2021adversarial,naseer2021intriguing}. Transformers have also been recently investigated for image segmentation~\cite{zheng2021rethinking,chen2021transunet,cao2021swin}. TransUNet~\cite{chen2021transunet} is the first Transformer based approach for medical image segmentation. It adapts a UNet structure, and replaces the bottleneck layer with ViT~\cite{dosovitskiy2020image} where patch embedding is applied on a feature map generated from CNN encoder (where input is a 2D slice of 3D volume). Unlike these hybrid approaches (using both convolutions and self-attention), Cao \etal~\cite{cao2021swin} proposed Swin-UNet, a purely transformer based network for medical image segmentation. It inherits swin-transformer blocks~\cite{liu2021swin} and shows better segmentation results  over TransUNet~\cite{chen2021transunet}. The 3D version of TransUnet~\cite{chen2021transunet}, called TransBTS~\cite{wang2021transbts} has a CNN encoder-decoder design and a Transformer as the bottleneck layer. Zhou \etal~\cite{zhou2021nnformer} proposed nnFormer with 3D Swin Transformer based blocks as encoder and decoder with interleaved stem of convolutions. A model which employs a transformer as the encoder and directly connects intermediate encoder outputs to the the decoder via skip connections is proposed in~\cite{hatamizadeh2022unetr}. The encoder-decoder structural comparison of SOTA methods are shown in~\cref{fig:architecture} (d).
The aforementioned transformer based approaches for 3D medical image segmentation have shown their promises, by achieving better performances compared with their CNN counterparts. Our proposed model, on the other hand, processes the volumetric data in its entirety, thus fully encoding the interactions between slices. Moreover, our proposed model is built purely based on Transformers and introduces lateral connections to perform CA along-with SA in the encoder-decoder design. These design elements contributed in achieving better segmentation performance, along-with enhanced robustness.

\section{Experiments}
\label{sec:exp}
\begin{minipage}[c]{0.6\textwidth}
\centering
    \resizebox{1\textwidth}{!}{
    \begin{tabular}{l|cc|cc|cc|cc}
    \toprule
       \multirow{2}{*}{Method}  & \multicolumn{2}{c|}{Average} & \multicolumn{2}{c|}{WT} & \multicolumn{2}{c|}{ET} & \multicolumn{2}{c}{TC}\\  \cline{2-9}
       &  HD95 $\downarrow$ & DSC $\uparrow$ & HD95 $\downarrow$ & DSC $\uparrow$ & HD95 $\downarrow$ & DSC $\uparrow$ & HD95 $\downarrow$ & DSC $\uparrow$ \\
       \hline
       \hline
       UNet~\cite{ronneberger2015u} & 10.19 & 66.4 & 9.21 & 76.6 & 11.12 & 56.1 & 10.24 & 66.5 \\
       AttUNet~\cite{oktay2018attention} & 9.97 & 66.4 & 9.00 & 76.7 & 10.45 & 54.3 & 10.46 & 68.3 \\
       nnUNet~\cite{isensee2021nnu} & 4.60 & 81.9 & \underline{3.64} & \textbf{91.9} & 4.06 & 80.97 & 4.91 & 85.35\\
       \hline
       SETR NUP~\cite{zheng2021rethinking} & 13.78 & 63.7 & 14.419 & 69.7 & 11.72 & 54.4 & 15.19 & 66.9 \\
       SETR PUP~\cite{zheng2021rethinking} & 14.01 & 63.8 & 15.245 & 69.6 & 11.76 & 54.9 & 15.023 & 67.0 \\
       SETR MLA~\cite{zheng2021rethinking} & 13.49 & 63.9 & 15.503 & 69.8 & 10.24 & 55.4 & 14.72 & 66.5 \\
       TransUNet~\cite{chen2021transunet} & 12.98 & 64.4 & 14.03 & 70.6 & 10.42 & 54.2 & 14.5 & 68.4 \\
       TransBTS~\cite{wang2021transbts} & 9.65 & 69.6 & 10.03 & 77.9 & 9.97 & 57.4 & 8.95 & 73.5 \\
       CoTr~\cite{xie2021cotr} & 9.70 & 68.3 & 9.20 & 74.6 & 9.45 & 55.7 & 10.45 & 74.8 \\
       UNETR~\cite{hatamizadeh2022unetr} & 8.82 & 71.1 & 8.27 & 78.9 & 9.35 & 58.5 & 8.85 & 76.1 \\
       nnFormer~\cite{zhou2021nnformer} & 4.05 & \underline{86.4} & 3.80 & 91.3 & 3.87 & \underline{81.8} & \underline{4.49} & \underline{86.0}\\
       \hline
       VT-UNet-S & \underline{3.84} & 85.9 & 4.01 & 90.8 & \underline{2.91} & \underline{81.8} & 4.60 & 85.0 \\
       VT-UNet-B & \textbf{3.43} & \textbf{87.1} & \textbf{3.51} & \textbf{91.9} & \textbf{2.68} & \textbf{82.2} &\textbf{4.10} & \textbf{87.2} \\
    \bottomrule
    \end{tabular}
    }
\captionof{table}{Segmentation Results on MSD BraTS Dataset.}
\label{tab:quantitative_results_1}
\end{minipage}
\begin{minipage}[c]{0.35\textwidth}
\centering
    \resizebox{1\textwidth}{!}{
    \begin{tabular}{l lcc}
    \toprule
       &Artefact~ & Avg. HD95 $\downarrow$ & Avg. DSC $\uparrow$  \\
       \hline
       \hline
       \parbox[t]{20mm}{\multirow{3}{*}{\makecell{nnFormer}}}
       & Clean & 4.05 & 86.4 \\
       & Motion & 4.81 & 84.3 \\
       & Ghost & 4.30 & 84.5 \\
       & Spike & 4.63 & 84.9  \\
       \hline
       \parbox[t]{20mm}{\multirow{3}{*}{\makecell{VT-UNet}}} 
       & Clean & 3.43 & 87.1 \\
       & Motion & 3.87 & 85.8 \\
       & Ghost & 3.69 & 86.0 \\
       & Spike & 3.50 & 86.6 \\
    \bottomrule
    \end{tabular}
    }
\captionof{table}{Robustness Analysis.}
    \label{fig:robustness}
    \vspace{2mm}
    \resizebox{0.95\textwidth}{!}{
    \begin{tabular}{l cc}
    \toprule
       VT-UNet-B~ & Avg. HD95 $\downarrow$ & Avg. DSC $\uparrow$  \\
       \hline
       \hline
       w/o FPE & 85.35 & 4.33 \\
       w/o FPE \& CA~~ & 83.58 & 6.18 \\
    \bottomrule
    \end{tabular}
    }
\captionof{table}{Ablation Study.}
    \label{fig:ablation}
\end{minipage}

\noindent \textbf{Implementation Details.}
We use 484 MRI scans from MSD BraTS task~\cite{antonelli2021medical}. Following~\cite{hatamizadeh2022unetr,zhou2021nnformer}, we divide 484 scans into 80\%, 15\% and 5\% for training, validation and testing sets, respectively. 
We use PyTorch~\cite{paszke2017automatic}, with a single Nvidia A40 GPU. The weights of Swin-T~\cite{liu2021swin} pre-trained on ImageNet-22K are used to initialize the model. For training, we employ AdamW optimizer with a learning rate of $1e^{-4}$ for 1000 epochs and a batch size of 4. We used rotating, adding noise, blurring and adding gamma as data augmentation techniques.
\begin{wraptable}{r}{6.2cm}
\tabcolsep=0.04cm
\scriptsize
\centering
\begin{tabular}{*{5}{c}}
\scriptsize{GT} & \scriptsize{VT-UNet} & \scriptsize{nnUNet} & \scriptsize{UNETR} & \scriptsize{nnFormer} \\
\includegraphics[width=0.185\linewidth, trim={3cm 2.5cm 3cm 3.5cm},clip]{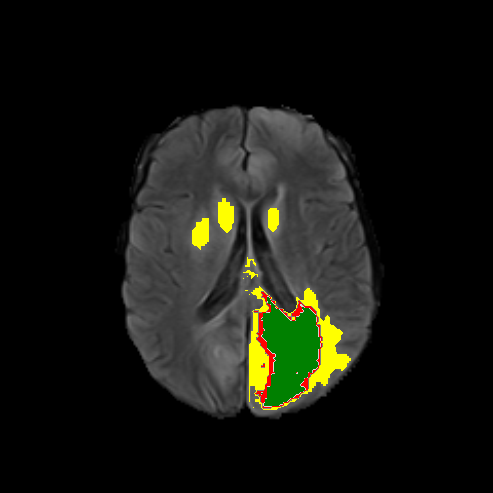} &
\includegraphics[width=0.185\linewidth, trim={3cm 2.5cm 3cm 3.5cm},clip]{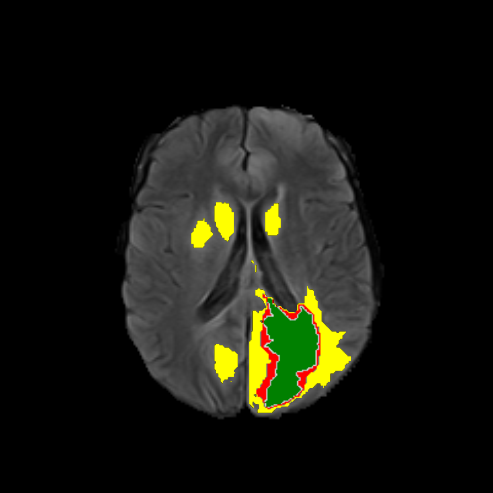} &
\includegraphics[width=0.185\linewidth, trim={3cm 2.5cm 3cm 3.5cm},clip]{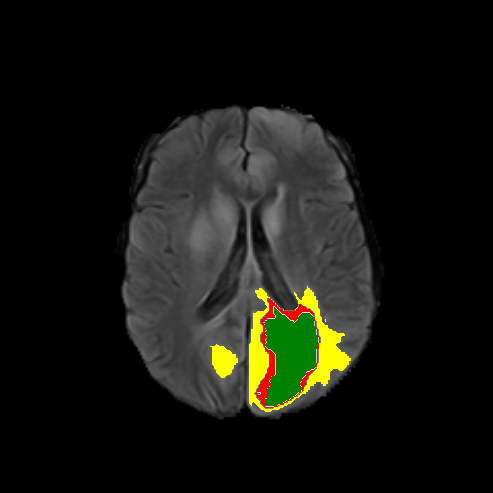} &
\includegraphics[width=0.185\linewidth, trim={3cm 2.5cm 3cm 3.5cm},clip]{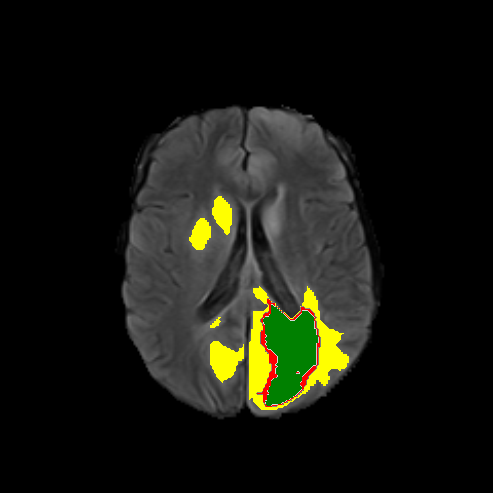} & 
\includegraphics[width=0.185\linewidth, trim={3cm 2.5cm 3cm 3.5cm},clip]{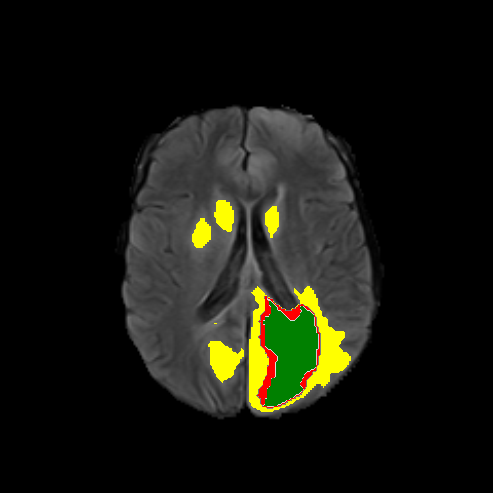}\\
\includegraphics[width=0.185\linewidth, trim={9cm 5cm 9cm 5cm},clip]{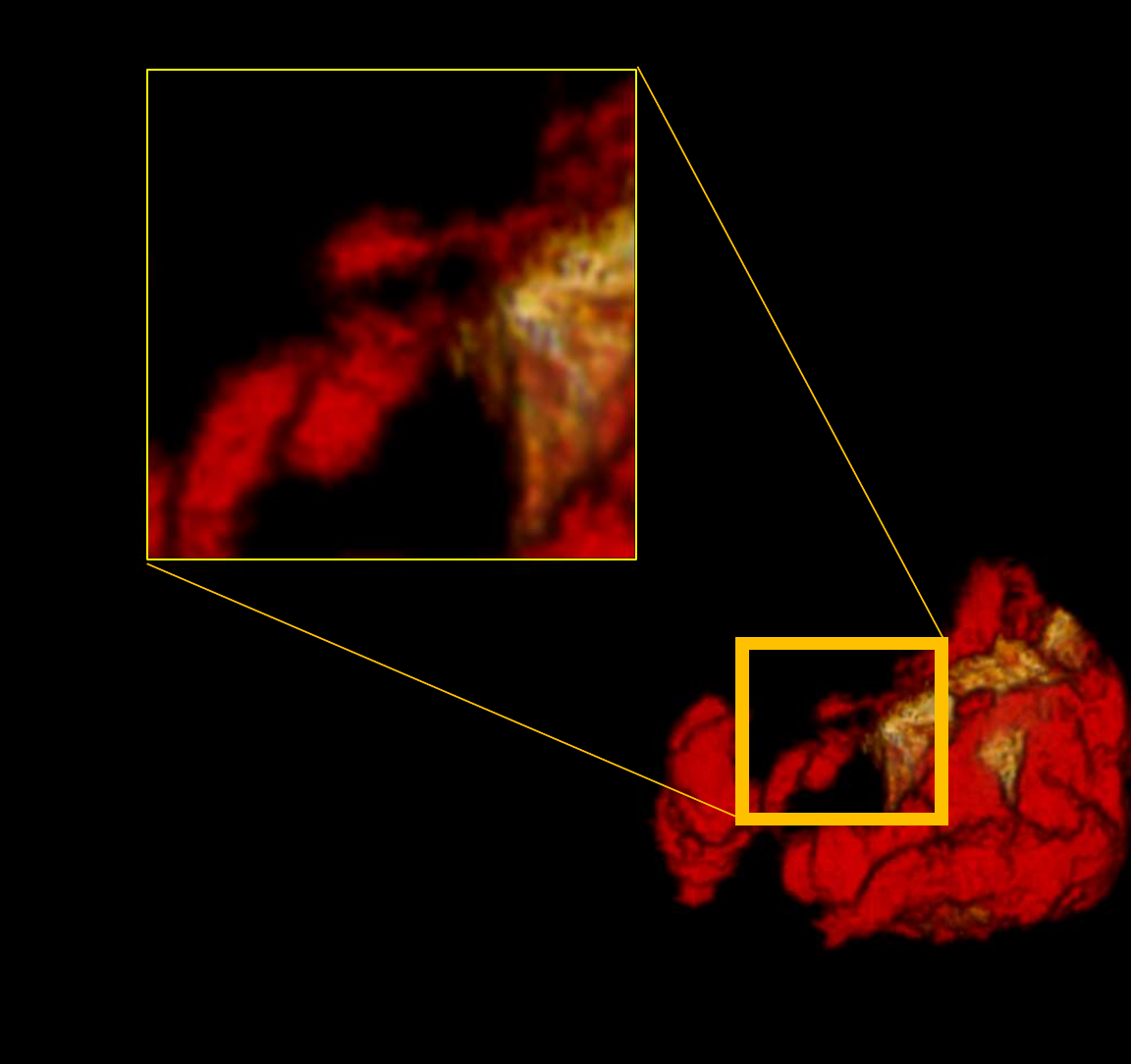} &
\includegraphics[width=0.185\linewidth, trim={9cm 5cm 9cm 5cm},clip]{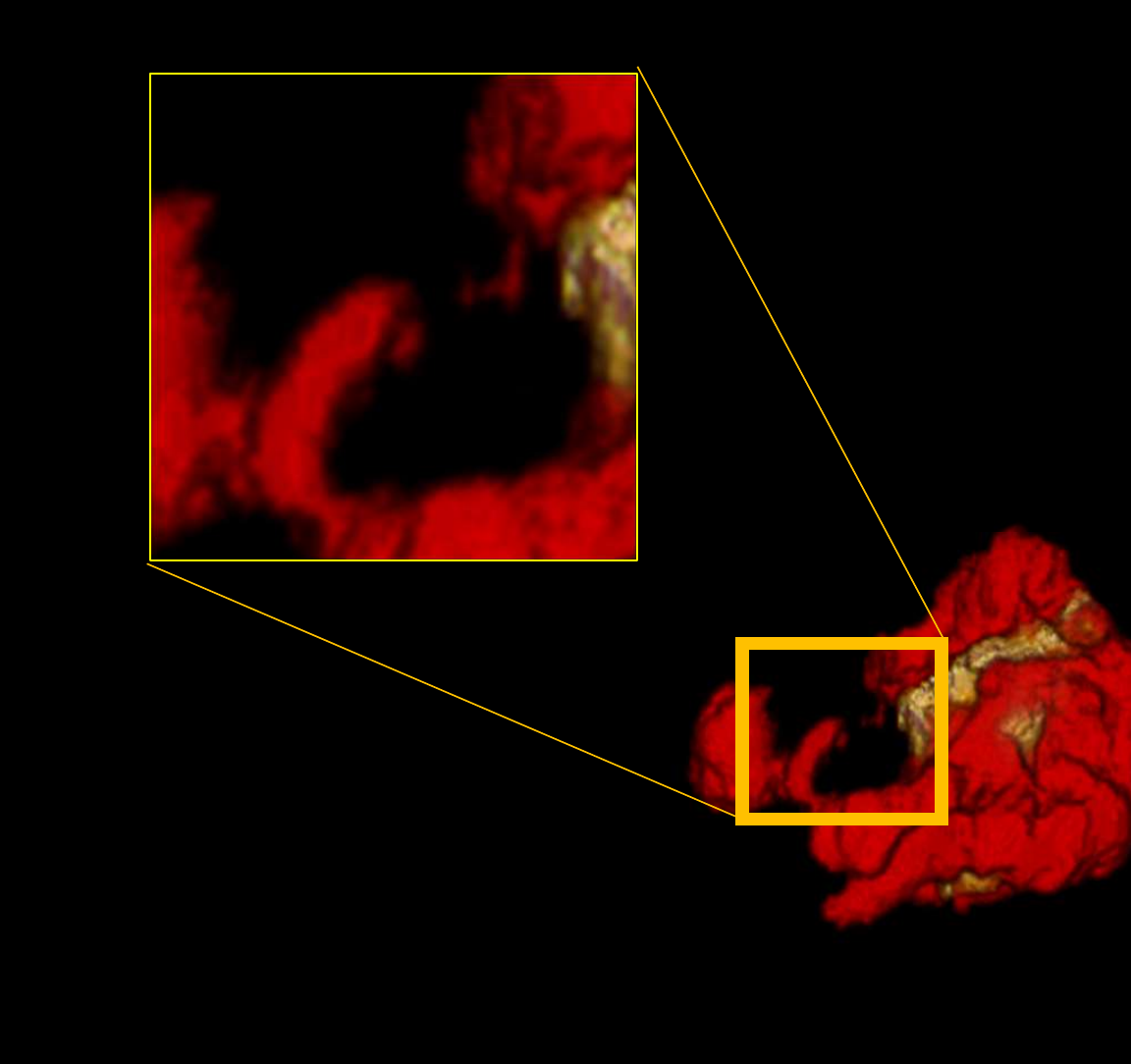} &
\includegraphics[width=0.185\linewidth, trim={9cm 5cm 9cm 5cm},clip]{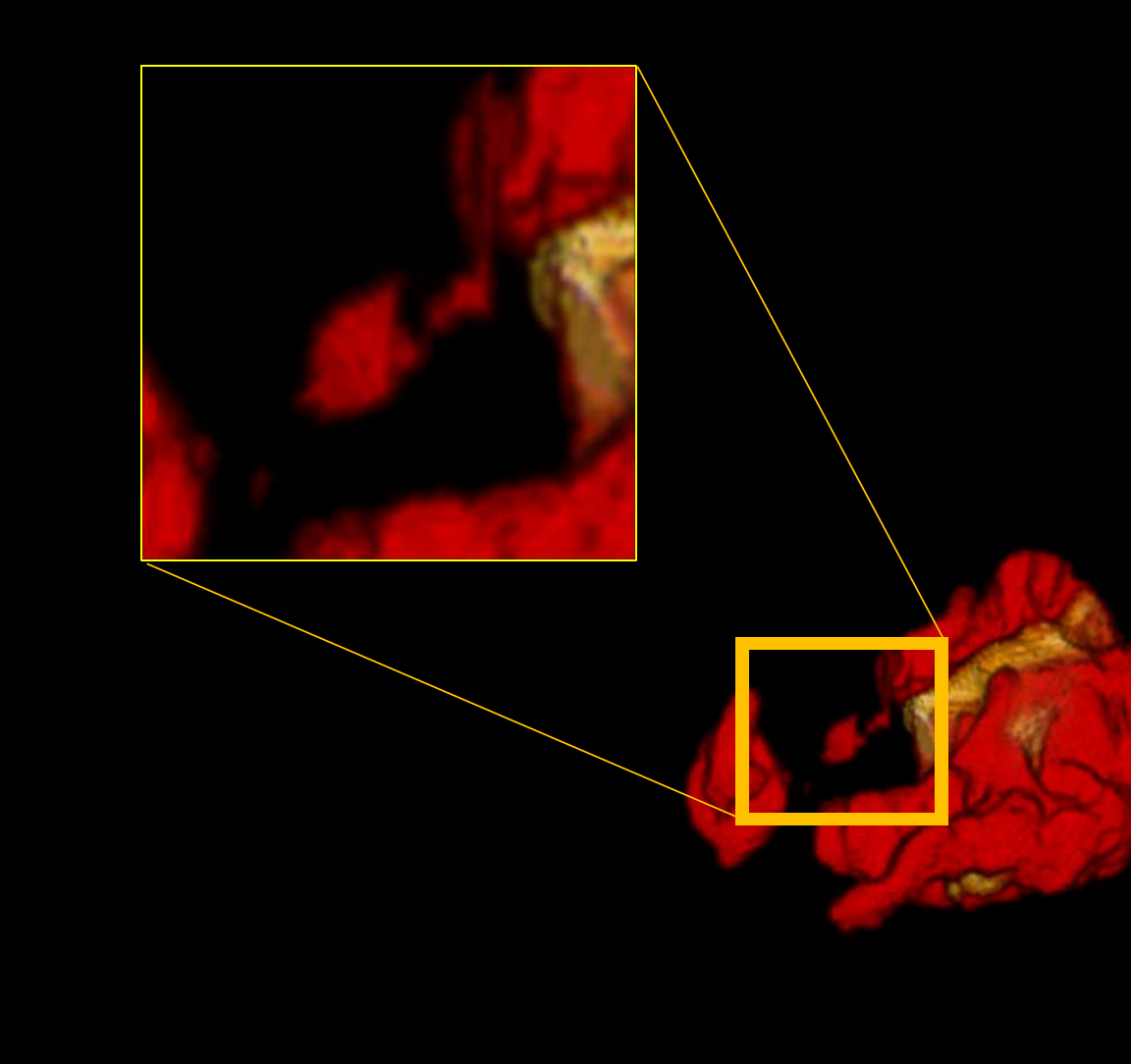} &
\includegraphics[width=0.185\linewidth, trim={9cm 5cm 9cm 5cm},clip]{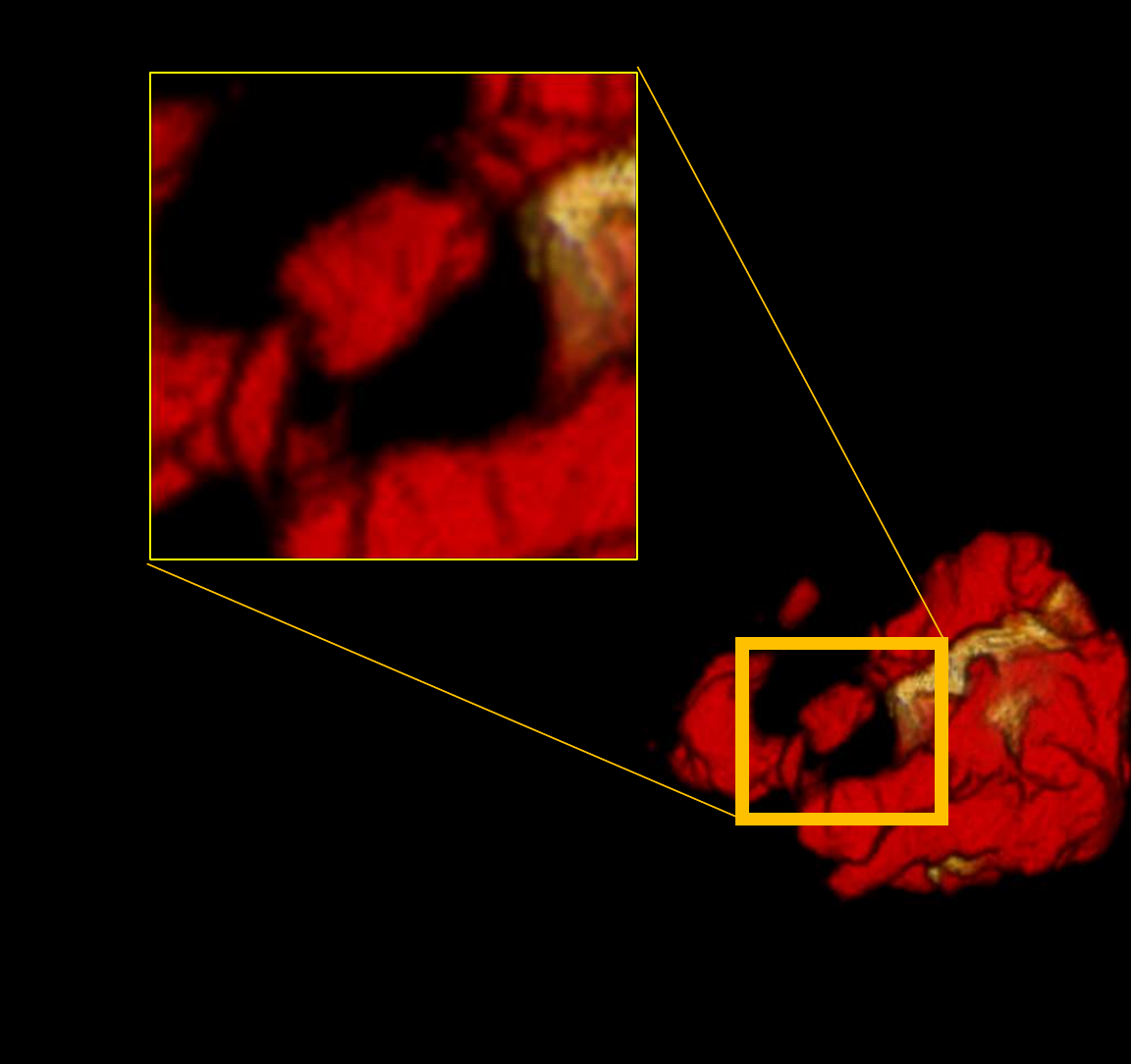} &
\includegraphics[width=0.185\linewidth, trim={9cm 5cm 9cm 5cm},clip]{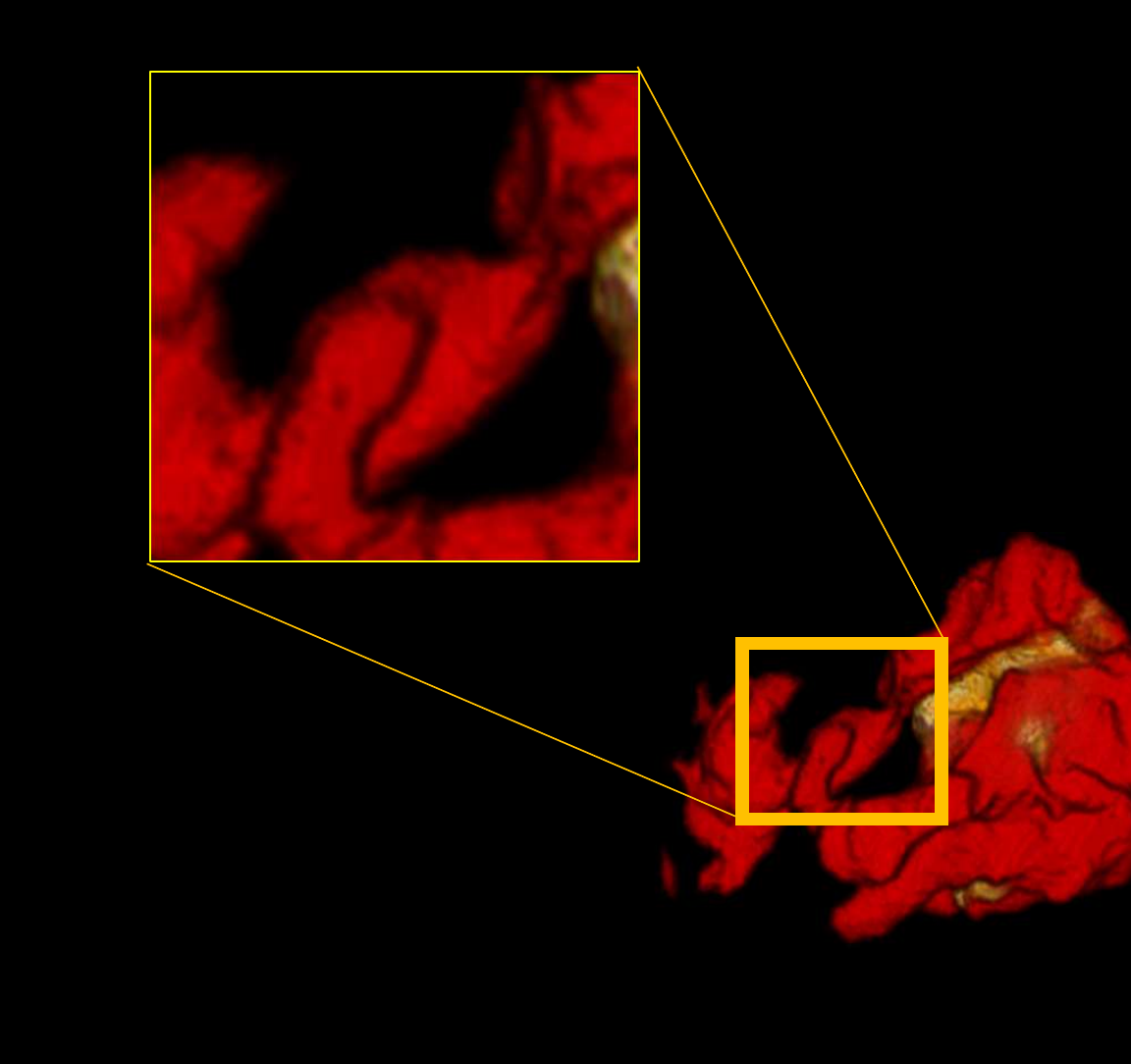} \\
\includegraphics[width=0.185\linewidth, trim={3cm 2.5cm 3cm 3.5cm},clip]{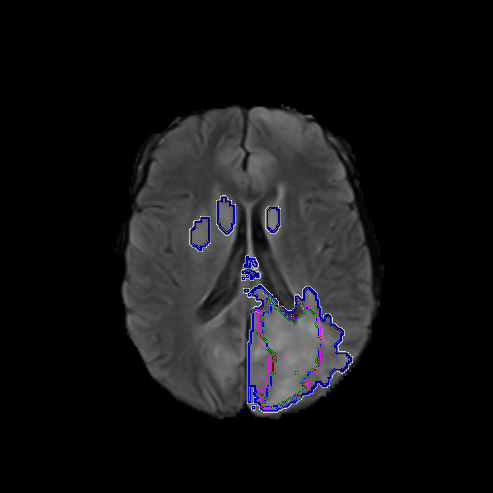} &
\includegraphics[width=0.185\linewidth, trim={3cm 2.5cm 3cm 3.5cm},clip]{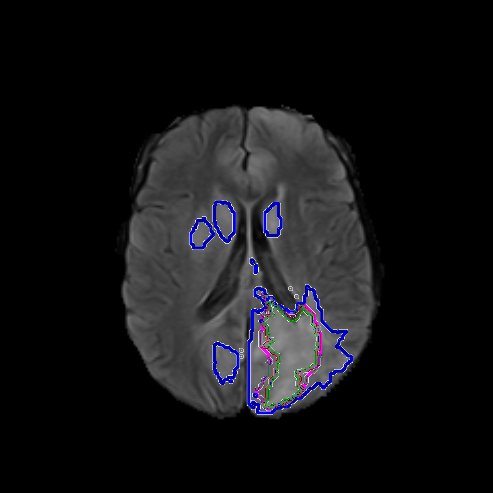} &
\includegraphics[width=0.185\linewidth, trim={3cm 2.5cm 3cm 3.5cm},clip]{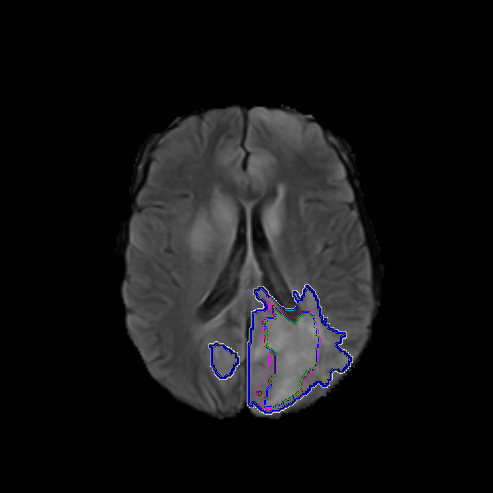} &
\includegraphics[width=0.185\linewidth, trim={3cm 2.5cm 3cm 3.5cm},clip]{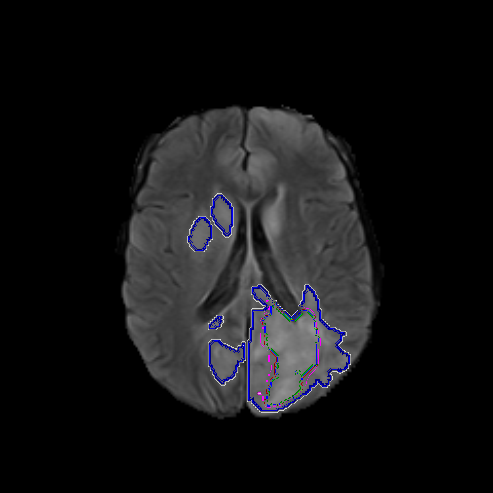} & 
\includegraphics[width=0.185\linewidth, trim={3cm 2.5cm 3cm 3.5cm},clip]{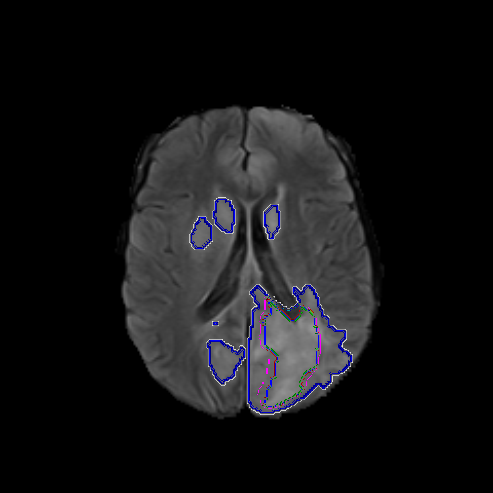}\\
\end{tabular}
\caption{Qualitative Segmentation Results. \textit{\textbf{Row-1}}: Yellow, Green and Red represent Peritumoral Edema  (ED), Enhancing Tumor (ET) and Non Enhancing Tumor (NET)/  Necrotic Tumor (NCR), respectively. \textit{\textbf{Row-2}}: volumetric tumor prediction. \textit{\textbf{Row-3}}: segmentation boundaries.}
\label{fig:qualitative_results}
\vspace{-6mm}
\end{wraptable}

\noindent \textbf{Experimental Results.} 
~\cref{tab:quantitative_results_1} compares VT-UNet with recent transformer based approaches and SOTA CNN based methods. We introduce variants of the VT-
UNet, by changing the number of embedded dimensions
used for model training. Our variants are:
\textbf{(a)} Small \emph{VT-UNet-S} with $C = 48$ \textbf{(b)} Base \emph{VT-UNet-B} with $C = 72$. We use Dice S$\o$rensen coefficient (DSC) and Hausdorff Distance (HD) as evaluation metrics, and separately compute them for three classes:
(1) Enhancing Tumor (ET), (2) the Tumor Core (TC) (addition of ET, NET and NCR), and (3) the Whole
Tumor (WT) (addition of ED to TC), following similar evaluation strategy as in~\cite{isensee2021nnu,hatamizadeh2022unetr,zhou2021nnformer}.  
Our quantitative results in \cref{tab:quantitative_results_1} suggest that VT-UNet achieves best overall performance in DSC and HD.~\cref{fig:qualitative_results} shows qualitative segmentation results on unseen patient data. We can observe that our model can accurately segment the structure and delineates boundaries of tumor. We believe that, capturing long-range dependencies across adjacent slices plays a vital role in our model's performance. 
Our empirical results in~\cref{fig:ablation} reveal the importance of introducing Parallel CA and SA together with FPE in VT-Dec-Blks along with convex combination. We can notice that all of these components contribute towards model's performance. 

\noindent \textbf{Robustness Analysis.~} Factors such as patient's movement and acquisition conditions can introduce noise to MRI. Here, we investigate the robustness of VT-UNet, by synthetically introducing  artefacts to MR images at inference time. These include \textbf{(1)} Motion artefacts~\cite{shaw2018mri}. \textbf{(2)} Ghosting artefacts~\cite{axel1986respiratory}. \textbf{(3)} Spike artefacts (Herringbone artefact)~\cite{jin2017mri}. ~\cref{fig:robustness} compares the robustness of our method with nnFormer~\cite{zhou2021nnformer}. The results suggest that VT-UNet performs more reliably in the presence of these nuisances. Our findings are consistent with existing works on RGB images, where Transformer based models have shown better robustness against occlusions \cite{naseer2021intriguing}, natural and adversarial perturbations \cite{shao2021adversarial,naseer2021intriguing}, owing to their highly dynamic and flexible receptive field.

\section{Conclusion}
This paper presents a volumetric transformer network for medical image segmentation, that is computationally efficient to handle large-sized 3D volumes, and learns representations that are robust against artefacts. Our results show that the proposed model achieves consistent improvements over existing state-of-the-art methods in volumetric segmentation. We believe our work can assist better clinical diagnosis and treatment planning. 
%
%
\bibliographystyle{splncs04}
\bibliography{references}

\clearpage
\section{Supplementary Material}

\paragraph{Preliminaries}
Here we provide a brief description about the SA. In SA, we are interested  in generating a new sequence of tokens, $\{\vec{z}_1,\vec{z}_2, \cdots ,\vec{z}_\tau\}$, from $\mathds{X}$ to better represent our signal according to the objective of learning. 
In doing so, we can assume that $\vec{x}_i$s span a subset of $\mathbb{R}^C$ and define $\vec{z}_i$ as a point in that span, \ie $\vec{z}_i = \sum_j a_{ij} \vec{x}_j$, where $a_{ij}$s are combination values defined by  $\vec{x}_i$ and $\vec{x}_j$. A possible choice for $a_{ij}$ is based on the similarity of $\vec{x}_i$ and $\vec{x}_j$, which algebraically is proportional to $\langle \vec{x}_i, \vec{x}_j\rangle$. Such a choice enables us to define the token $\vec{z}_i$ by attending to important parts of the input sequence according to the objective in hand, hence the name attention. 
We can take a further step and put a constraint on our design by enforcing the generated tokens to lie inside the convex hull defined by  $\vec{x}_i$s. In that case, we will have $\vec{z}_i = \sum_j a_{ij} \vec{x}_j, a_{ij} \geq 0, \sum_j a_{ij} = 1$. The convex hull formulation endows nice properties, one being that the resulting tokens cannot grow boundlessly, given the fact that the input is assumed to be a natural signal. The SA operation is built upon the above idea with some modifications. Firstly, we assume that tokens, in their original form,  might not be optimal for defining the span. Therefore, in SA we define the span by learning a linear mapping from the input tokens. This we can show with $\mathbb{R}^{\tau \times C_v} \ni \vec{V} = \vec{X}\vec{W}_\vec{V}$, where we stack tokens $\vec{x}_i$s into the rows of $\vec{X}$ (\ie, $\vec{X}=[\vec{x}_1|\vec{x}_2|\cdots\vec{x}_\tau]^\top$). Then we turn our attention to $a_{ij}$ and define it by learning two linear mappings, following a similar argument. In particular, first we define a set of keys from  $\vec{x}_i$ as $\mathbb{R}^{\tau \times C_k} \ni \vec{K} = \vec{X}\vec{W}_\vec{K}$. To generate $a_{ij}$, we measure the similarity of a transformed version of $\vec{x}_i$, which we call the query $\vec{q}_i = \vec{W}_\vec{Q}^\top\vec{x}_i$, with respect to the keys $\vec{k}_j$ that are represented by the rows of $\vec{K}$. That is, $a_{ij} \propto \langle \vec{q}_i,\vec{k}_j\rangle$. Put everything into a matrix form and opt for a softmax function to achieve the constraints $a_{ij} \geq 0$ and $\sum_j a_{ij} = 1$, we arrive at Eq (1).

\paragraph{Breaking Symmetry Cont.}
The linear patch-projection flattens the features, thereby failing to fully encapsulate object-aware representations (e.g., spatial information) that are critical for anatomical pixel-wise segmentation. As shown in Fig. 2(e), combining two sets of tokens may results in loss of fluency. Therefore, in order to preserve features among continuous slices, we supplement the 3D Fourier Feature Positional Encoding (FPE) for the tokens generated from MSA module by adapting sine and cosine functions with different frequencies~\cite{vaswani2017attention}, which provides unique encoding for each input token. Following work by Wang \etal~\cite{wang2021translating}, we used the extended version of 2D positional encoding for 3D. Therefore, we call this as a 3D FPE or in other words a 3D positional encoding with a sinusoidal input mapping for VT-Dec-Blks.  After applying 3D FPE to tokens, we pass it through a LN and MLP layer. Our empirical evaluations confirm that adding a Fourier feature positional bias can improve the poor conditioning of the feature representation. 

\paragraph{Loss Function.}
To train VT-UNet, we jointly minimize the Dice Loss together with Cross Entropy loss (computed in a voxel-wise manner). 

\end{document}